# pISTA-SENSE-ResNet for Parallel MRI Reconstruction

Tieyuan Lu, Xinlin Zhang, Yihui Huang, Yonggui Yang, Gang Guo, Lijun Bao, Feng Huang, Di Guo, Xiaobo Qu

*Abstract*—Magnetic resonance imaging has been widely applied in clinical diagnosis, however, is limited by its long data acquisition time. Although imaging can be accelerated by sparse sampling and parallel imaging, achieving promising reconstruction images with a fast reconstruction speed remains a challenge. Recently, deep learning approaches have attracted a lot of attention for its encouraging reconstruction results but without a proper interpretability. In this letter, to enable high-quality image reconstruction for the parallel magnetic resonance imaging, we design the network structure from the perspective of sparse iterative reconstruction and enhance it with the residual structure. The experimental results of a public knee dataset show that compared with the optimization-based method and the latest deep learning parallel imaging methods, the proposed network has less error in reconstruction and is more stable under different acceleration factors.

## I. INTRODUCTION

Magnetic resonance imaging (MRI) plays an indispensable role in clinical diagnosis [1]. Parallel imaging [2] and sparse sampling [1] have been proposed to speed up the MRI signal acquiring process by reducing the acquisition of k-space data. For the sparse sampling, many optimized-based methods have been proposed to reconstruct the magnetic resonance (MR) images under the compressed sensing (CS) theory. Reconstruction results of high quality can be acquired at the assumption that images are self-sparse or sparse in other domains [3]. However, the time-consuming iterations and the challenge to choose the optimal transform hinder its wide application. For parallel MRI, the traditional reconstruction methods include sensitivity encoding (SENSE) [4], and etc. Even the parallel MRI has been widely adopted as the acceleration scan, the accelerating factor is still limited under the number of coils in theoretically [2]. Therefore, besides applying the two approaches individually, the combinations of them are also proposed to earn a lager accelerating factor [5]. A typical combination is SparseSENSE [5], which applies sparse sampling to SENSE directly.

To achieve the lower reconstruction error by a sparser representation, many data-free sparse transforms have been used including the fixed ones [1, 6, 7] and the adaptive ones [3, 8, 9]. Recently, the data-based deep learning network has been applied in many applications including biological magnetic spectroscopy [10], medical image analysis [11], and accelerating MR images reconstruction [12-16], and obtained outstanding performances by the powerful expression ability of deep convolution neural network (CNN). However, compared with optimization-based iterative algorithms in MR image reconstruction, CNN appears to be a black box to the image reconstruction process. Meanwhile, the absence of guide line for constructing the structure of networks hinders proper usage of CNN in specific application. Therefore, some optimization-based algorithms have been unrolled and yielded as deep networks, such as ADMM-Net [17], ISTA-Net [18] and other variations [19], which outperform the optimization-based CS methods by large margins. However, none of the above networks can work for parallel MRI directly.

Within the field of fast MRI, Liu et.al proposed the projected iterative soft-thresholding algorithm (pISTA) [20] which can recover superior image quality and take lower memory consumption compared with other iterative methods, such as iterative soft-thresholding algorithm (ISTA) [21]. Ting et.al independently proposed the balanced sparse reconstruction method, which can solve the sparseSENSE model, called bFISTA-SENSE in [22]. Here we call it pFISTA-SENSE. Zhang et.al theoretically proved the convergence of pFISTA-SENSE [23].

In this letter, inspired by unrolling iterative reconstructions

This work was supported in part by National Key R&D Program of China (2017YFC0108703), National Natural Science Foundation of China (61571380, 61971361, 61871341, and 61811530021), Natural Science Foundation of Fujian Province of China (2018J06018), Fundamental Research Funds for the Central Universities (20720180056), Science and Technology Program of Xiamen (3502Z20183053), and China Scholarship Council. The authors would like thank the GPU donated by NIVDIA Corporation. (*Corresponding author*: Xiaobo Qu)

Tieyuan Lu, Xinlin Zhang, Yihui Huang, Lijun Bao, and Xiaobo Qu are with the Department of Electronic Science, Fujian Provincial Key Laboratory of Plasma and Magnetic Resonance, School of Electronic Science and Engineering (National Model Microelectronics College), Xiamen University, Xiamen, China (e-mail: tieyuanlu@stu.xmu.edu.cn; xinlin@stu.xmu.edu.cn; huangyihui@stu.xmu.edu.cn; baolijun@xmu.edu.cn; quxiaobo@xmu.edu.cn).

Yonggui Yang and Gang Guo are with the Department of Radiology, the Second Affiliated Hospital of Xiamen Medical College, Xiamen, China (e-mail: yangyonggui125@sina.cn; guogangxm@163.com)

Feng Huang is with Neusoft Medical System, Shanghai, China (e-mail: ethan@neusoft.com).

Di Guo is with the School of Computer and Information Engineering, Fujian Provincial University Key Laboratory of Internet of Things Application Technology, Xiamen University of Technology, Xiamen, China (e-mail: guodi@xmut.edu.cn).

as the structure of networks, we design the unrolled network by replacing the hand-craft transform with CNN properly and use the residual structure to enhance the performance, expecting to get a lower reconstruction error than the existing method. In our experiments of a public knee dataset [24], the proposed network achieves the best reconstruction results compared with the state-of-the-art methods under different acceleration factors.

## II. PROPOSED METHOD

The spaseSENSE [5] model, which applies sparse sampling to SENSE directly, can be written as:

$$\min_{\mathbf{x}} \lambda \|\mathbf{\Psi x}\|_1 + \frac{1}{2}\sum_{j=1}^{J}\|\mathbf{UFC}_j\mathbf{x} - \mathbf{y}_j\|_2^2, \quad (1)$$

where $\mathbf{x}$ is the desired composite MR image rearranged to a column vector, $\lambda$ is the regularization parameter, $\mathbf{\Psi}$ is the sparse transform which transposes the image into the space of sparse coefficients, $\mathbf{C}_j$ is the sensitivity map of $j^{th}$ coil, $\mathbf{F}$ is the Fourier transform, $\mathbf{U}$ is the undersampling matrix, $\mathbf{y}_j$ is the acquired undersampled k-space data of the $j^{th}$ coil.

In our proposed network, we unrolled pFISTA-SENSE [20, 22, 23] as the basic structure of our network. The pFISTA-SENSE solves the reconstruction problem in equation (1) iteratively and the $s^{th}$ iteration can be written as following:

$$\begin{cases} \mathbf{t}_s = \mathbf{x}_s + \gamma \sum_{j=1}^{J}\mathbf{C}_j^H\mathbf{F}^H\mathbf{U}^T\left(\mathbf{y}_j - \mathbf{UFC}_j\mathbf{x}_s\right) \\ \boldsymbol{\alpha}_s = \mathbf{\Psi t}_s \\ \tilde{\boldsymbol{\alpha}}_{s+1} = T_{\lambda\gamma}\left(\boldsymbol{\alpha}_s\right) \\ \mathbf{x}_{s+1} = \mathbf{\Phi}\tilde{\boldsymbol{\alpha}}_{s+1}, \end{cases} \quad (2)$$

where $\gamma$ is the step size, the superscript $H$ and $T$ denote conjugate transpose and transpose respectively, $T_{\lambda\gamma}$ is a point wise soft-thresholding function:

$$T_{\lambda\gamma}(\beta) = \max\{|\beta| - \gamma\lambda, 0\} \cdot \frac{\beta}{|\beta|} \quad (3)$$

$\boldsymbol{\alpha}$ is the sparse coefficients of $\mathbf{x}$ under the forward transform $\mathbf{\Psi}$. And $\mathbf{\Phi}$ is the backward transform corresponding to $\mathbf{\Psi}$.

Specifically, the four update steps in equation (2) correspond to four different operations or modules of an iteration block in our proposed pISTA-SENSE-ResNet. However, the iteration blocks in network is fixed in a small number $S$ compared with the pISTA-SENSE.

### A. Data Consistency (DC) Module

The DC module in our pISTA-SENSE-ResNet keeps same with the one in pFISTA-SENSE, which is built to keep the data consistency between the acquired k-space $\mathbf{y}_j$ data and the predicted one $\mathbf{x}_s$ by the network and formulated as:

$$\mathbf{t}_s = \mathbf{x}_s + \gamma_s \sum_{j=1}^{J}\mathbf{C}_j^H\mathbf{F}^H\mathbf{U}^T\left(\mathbf{y}_j - \mathbf{UFC}_j\mathbf{x}_s\right) \quad (4)$$

### B. Forward Operation

In pFISTA-SENSE, the sparse coefficients of images under the sparse transform are required to be as sparse as possible. However, the hand-craft transform $\mathbf{\Psi}$ is difficult to tune for different kinds of image reconstruction.

Different from pFISTA-SENSE, here we replace $\mathbf{\Psi}$ with a learnable CNN, called forward operation $P_s$:

$$\boldsymbol{\alpha}_s = P_s\mathbf{t}_s \quad (5)$$

The forward operation $P_s$ contains $L$ convolution layers: $\mathbf{W}^{f,1},\ldots,\mathbf{W}^{f,L}$, and each layer contains $K$ filters whose size are $h \times h$. Besides, the non-linear activate function ReLU [25] is added behind all the convolution layers except for the last one.

### C. Soft-thresholding Module

The soft-thresholding module mimics the soft-thresholoding function in pFISTA-SENSE. What should be noted is that the threshold value $\gamma_s\lambda_s$ varies in different iteration blocks, while in pFISTA-SENSE, the thresh value $\gamma\lambda$ keeps same and fixed.

$$\tilde{\boldsymbol{\alpha}}_{s+1} = T_{\lambda_s\gamma_s}\left(\boldsymbol{\alpha}_s\right) \quad (6)$$

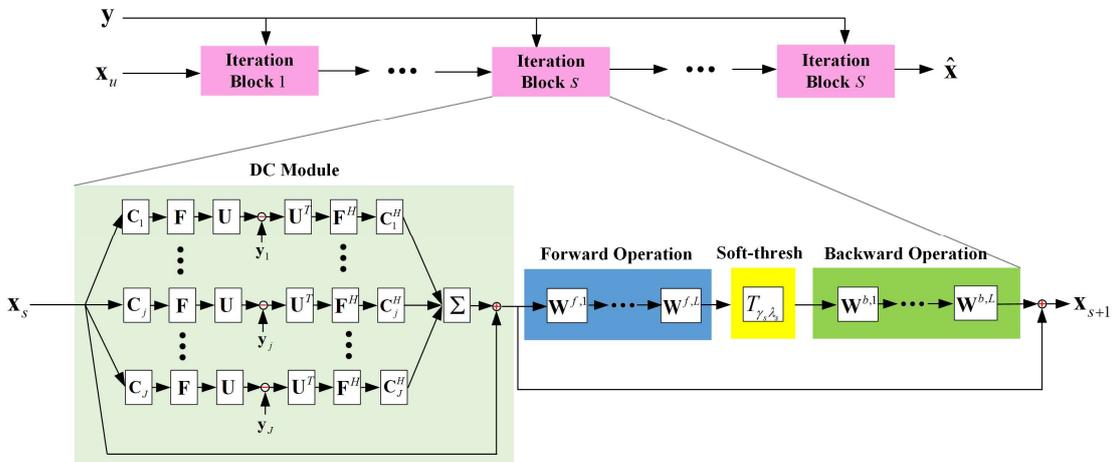

Fig.1. The proposed pISTA-SENSE-ResNet for parallel MRI reconstruction

*D. Backward Operation*

The CNN-based backward operation $Q_s$, which simulates the function of $\Phi$ in the optimization-based algorithm, has the same number of convolution layers as $P_s$. The convolution layers in $Q_s$ are noted by $\mathbf{W}^{b,1},\ldots,\mathbf{W}^{b,L}$, possessing the same number filters as $\mathbf{W}^{f,1},\ldots,\mathbf{W}^{f,L}$. The backward operation $Q_s$ transforms the thresholded coefficients $\tilde{\boldsymbol{\alpha}}_{s+1}$ to image domain:

$$\mathbf{x}_{s+1} = Q_s \tilde{\boldsymbol{\alpha}}_{s+1} \qquad (7)$$

And the network contains equation (7) is pISTA-SENSE-Net.

*E. The Residual in pISTA-SENSE-ResNet*

Inspired by the ResNet [26], we add a skip connection to sum the input of $P_s$ and the output of $Q_s$ in the iteration block to stabilize convergence. And the corresponding network is the proposed pISTA-SENSE-ResNet.

$$\mathbf{x}_{s+1} = \mathbf{t}_s + Q_s\left(T_{\lambda_s \gamma_s}\left(P_s \mathbf{t}_s\right)\right) \qquad (8)$$

*F. The Loss Function Design*

The loss function in pISTA-SENSE-ResNet is specified by the sum mean square error between the fully sampled coil-combined image $\mathbf{x}_n$ and the output of each iteration block $\mathbf{x}_{n,s}$, which has been used in [10]:

$$E = \sum_{n=1}^{N}\sum_{s=2}^{S+1} \left\| \mathbf{x}_n - \mathbf{x}_{n,s} \right\|_2^2, \qquad (9)$$

where $N$ is the number of MR images in the training dataset.

### III. Experimental Results

*A. Dataset and Network Setup*

The public knee dataset provided by [24] is used in our experiment. Sequence parameters of the knee data we used are: TR = 2870 ms, TE = 33 ms, in-plane resolution $0.49 \times 0.44$ mm$^2$ and a 15-element knee coil was equipped in scanning. We choose 10 persons as the training dataset and another 5 persons for testing, and each person contains 20 slices of size $320 \times 320$. Coil sensitivity maps are precomputed from a data block of size $24 \times 24$ at the center of k-space using ESPIRiT [27].

The 1D Cartesian sampling with random phase encoding is adopted here (shown in Fig.3. (f)), which emulates the part sampling of the phase encoding. The acceleration is realized by reducing sampling density along the phase encoding direction, while keeping fully sampled along the frequency direction according to the MRI sampling theory.

In the implementation of proposed pISTA-SENSE-ResNet, 10 iteration blocks are connected sequentially to build the whole network. Both the forward operation and the backward operation consist of 3 convolution layers, and each layer contains 48 filters whose size is $3 \times 3$. All filters are initialized using "Xavier" initialization [28], and Adam [29] is selected as the optimizer in the training progress with a learning rate of 0.001. $\gamma_s$ and $\lambda_s$ are initialized as 1 and 0.001 respectively.

*B. Results Comparison*

To evaluate the performance of the proposed and other methods in an objective view, we use the relative $l_2$ norm error (RLNE) and mean structure similarity index measure (MSSIM) [30] as the quantitative criteria. The RLNE is defined as:

$$\text{RLNE} = \left\| \mathbf{x} - \hat{\mathbf{x}} \right\|_2 / \left\| \mathbf{x} \right\|_2, \qquad (10)$$

where $\mathbf{x}$ denotes the fully sampled coil-combined image and $\hat{\mathbf{x}}$ represents the reconstructed image. The lower RLNE stands for the higher consistency between $\mathbf{x}$ and $\hat{\mathbf{x}}$. The MSSIM between $\mathbf{x}$ and $\hat{\mathbf{x}}$ is calculated by:

$$\text{MSSIM}(\mathbf{x}, \hat{\mathbf{x}}) = \frac{1}{M}\sum_{m=1}^{M} \frac{\left(2\mu_{\mathbf{x}_m}\mu_{\hat{\mathbf{x}}_m} + C_1\right)\left(2\sigma_{\mathbf{x}_m \hat{\mathbf{x}}_m} + C_2\right)}{\left(\mu_{\mathbf{x}_m}^2 + \mu_{\hat{\mathbf{x}}_m}^2 + C_1\right)\left(\sigma_{\mathbf{x}_m}^2 + \sigma_{\hat{\mathbf{x}}_m}^2 + C_2\right)} \qquad (11)$$

where $M$ is the number of local windows, $\mu_{\mathbf{x}_m}, \mu_{\hat{\mathbf{x}}_m}, \sigma_{\mathbf{x}_m}, \sigma_{\hat{\mathbf{x}}_m}$, and $\sigma_{\mathbf{x}_m \hat{\mathbf{x}}_m}$ denote the means, standard deviations, and covariance of the local window $\mathbf{x}_m$ and $\hat{\mathbf{x}}_m$. Constants $C_1$ and $C_2$ are added in case of the product of $\mu_{\mathbf{x}_m}^2 + \mu_{\hat{\mathbf{x}}_m}^2$ and $\sigma_{\mathbf{x}_m}^2 + \sigma_{\hat{\mathbf{x}}_m}^2$ close to zero. The higher MSSIM stands for the more excellent structure detail recovery.

We first compare the performance of the pISTA-SENSE-ResNet and pISTA-SENSE-Net at the acceleration factor (AF) of 7. As shown by the RLNE curve of training progress in Fig.2, residual structure indeed helps the pISTA-SENSE-ResNet converges more steadily and reaches a lower RLNE.

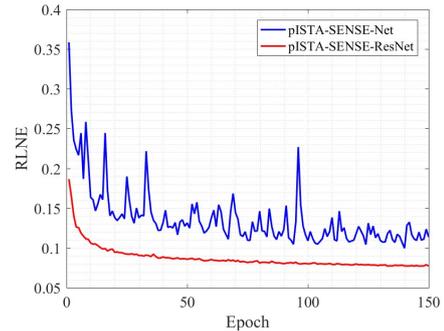

Fig. 2. The converge comparison between pISTA-SENSE-ResNet and pISTA-SENSE-Net on 1D sampling pattern (AF=7).

Comparing methods include the data-free pFISTA-SENSE [20, 23] and two data-based deep learning methods: VN [26] and MoDL [31], which work for the parallel MRI. For pFISTA-SENSE, the patch based directional wavelets [8] are utilized as sparse transforms, while the configurations of VN and MoDL keep default.

To show the stability of the proposed method under different acceleration factors, three sampling patterns with acceleration factors of 5, 7, and 9 are adopted. All the networks finish the training after 150 epochs. Among these reconstruction results, pISTA-SENSE-ResNet recovers the image more close to the fully sampled images: first, some details are recovered faithfully, while others not, such as the red arrow pointed in Fig.3 and Fig.4, second, the excessively sharp edge are well suppressed as the green arrows pointed in Fig.4. Besides this, less error is observed at the reconstruction results of pISTA-SENSE-ResNet according to the error maps.

As shown in the TABLE I, pISTA-SENSE-ResNet reaches the lowest mean value of RLNE, which means the average reconstruction error is the least among these methods. The corresponding standard deviation is comparable or a litter lower,

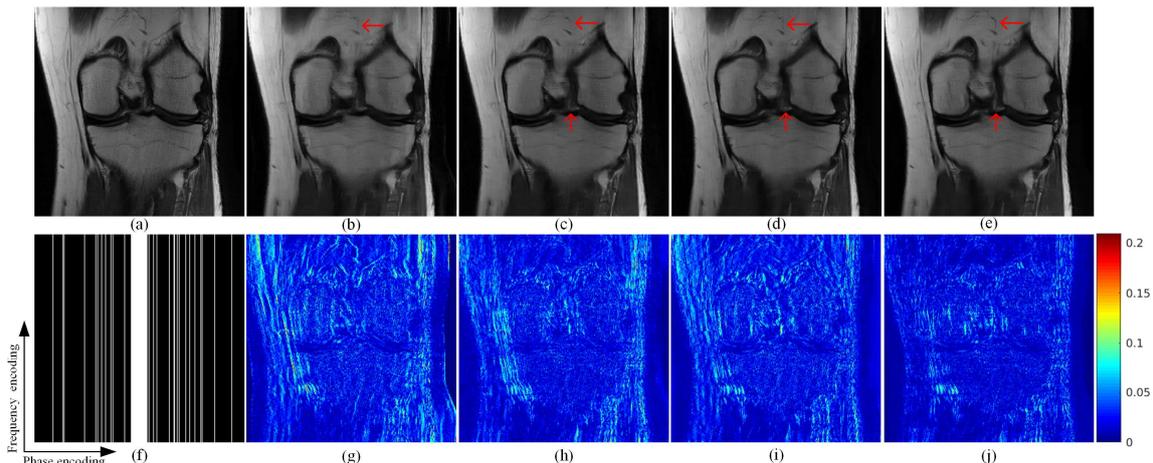

Fig. 3. Reconstruction results comparison (AF = 7): (a) the fully sampled coil-combined image, (b-e) the reconstruction of pFISTA-SENSE, VN, MoDL, and pISTA-SENSE-ResNet respectively, (f) is the sampling pattern, (g-j) are the error maps of (b-e), respectively.

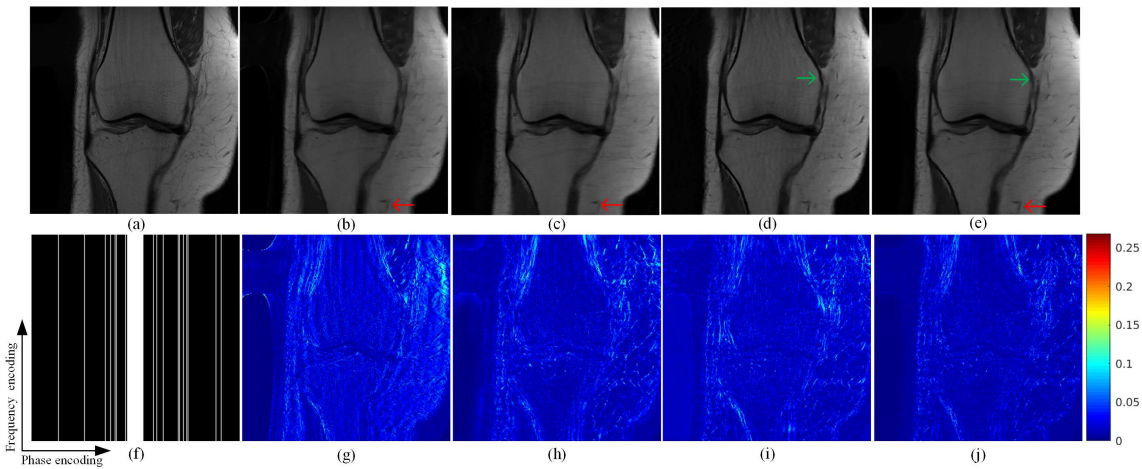

Fig. 4. Reconstruction results comparison (AF = 9): (a) the fully sampled coil-combined image, (b-e) the reconstruction of pFISTA-SENSE, VN, MoDL, and pISTA-SENSE-ResNet respectively, (f) is the sampling pattern, (g-j) are the error maps of (b-e), respectively.

TABLE I
OBJECTIVE CRITERIA ON THE TESTING DATASET

| Method | AF=5 | | AF=7 | | AF=9 | | Time CPU/GPU |
|---|---|---|---|---|---|---|---|
| | RLNE × 100 | MSSIM × 100 | RLNE × 100 | MSSIM × 100 | RLNE × 100 | MSSIM × 100 | |
| pFISTA-SENSE | 8.24 ± 1.14 | 88.65 ± 2.92 | 10.25 ± 1.34 | 84.08 ± 2.88 | 12.33 ± 4.69 | 81.30 ± 3.71 | 624s/- |
| VN | 7.11 ± 0.92 | 89.87 ± 2.16 | 9.87 ± 1.36 | 84.04 ± 2.54 | 11.37 ± 1.92 | 81.42 ± 2.78 | -/0.27s |
| MoDL | 6.73 ± 0.86 | 85.25 ± 6.38 | 9.05 ± 1.26 | 80.64 ± 6.03 | 10.13 ± 1.70 | 78.75 ± 4.65 | -/5.17s |
| pISTA-SENSE-ResNet | **6.29 ± 0.86** | **91.46 ± 2.20** | **8.68 ± 1.27** | **86.55 ± 2.46** | **9.78 ± 1.54** | **84.18 ± 2.81** | -/0.91s |

Note: The mean value and standard division are respect to the whole testing dataset under certain method and acceleration factor (AF). And the average reconstruction time per slice for each method is also attached in the last column.

which means the more steady reconstruction of the proposed. In terms to MSSIM, pISTA-SENSE-ResNet also gets the best in mean and standard deviation, except for the little higher standard deviation in AF = 5 and AF = 9, which demonstrates the superior detail recovery for the whole testing dataset.

## IV. CONCLUSION

In this letter, we proposed the pISTA-SENSE-ResNet to solve the parallel magnetic resonance imaging reconstruction problem, and the architecture of which is followed by unrolling the pFISTA-SENSE. Trainable filters are applied in the network, which get more adaptive to the specific magnetic resonance images and do not need to choose transforms manually. At the same time, the long iteration step in the pFISTA-SENSE is taken place by the cascade iteration blocks of the network, which favors the real-time reconstruction. Compared with the state-of-the-art deep learning methods under the public knee dataset, pISTA-SENSE-ResNet can recover images with more faithful details. In summary, these results show that our proposed network can recover magnetic resonance images of high quality in parallel magnetic resonance imaging reconstruction.